\documentclass[12pt]{article}
\usepackage{amsmath,amssymb}
\usepackage{latexsym,graphicx,epsfig}
\usepackage{curves,eepic}

\newcommand{\be}{\begin{equation}}
\newcommand{\ee}{\end{equation}}
\newcommand{\bes}{\begin{subequations}}
\newcommand{\ees}{\end{subequations}}
\newcommand{\bea}{\begin{eqnarray}}
\newcommand{\eea}{\end{eqnarray}}
\newcommand{\bear}{\begin{equation}\begin{array}}
\newcommand{\eear}[1]{\end{array}\label{#1}\end{equation}}
\def\ba{$$\begin{array}}
\def\ea{\end{array}$$}
\def\bra{$\begin{array}}
 \def\era{\end{array}$}

\newcommand{\fr}[2]{\dfrac{{ #1}}{{ #2}}}

\newcommand{\la}{\langle}
\newcommand{\ra}{\rangle}
\newcommand{\fn}[1]{\footnote{{\sf #1}}}

\newcommand{\epe}{\mbox{$e^+e^-\,$}}
\newcommand{\ggam}{\mbox{$\gamma\gamma\,$}}
\newcommand{\egam}{\mbox{$e\gamma \,$}}

\newcommand{\ggww}{\mbox{$\gamma\gamma\to W^+W^-\,$}}

\newcommand{\gewnu}{\mbox{$e\gamma\to \nu W\,$}}

\newcommand{\CP}{${\cal C}{\cal P}\;$}

\def\cl{\centerline}

\newsavebox{\fmbox}

{\end{list}}
\newcounter{enumct}

\newcommand{\bu}{$\bullet$\ }

\usepackage{wrapfig}

\begin{document}
\renewcommand{\tilde}{\widetilde}%[1]

\date{}
\title{Physical problems for future Photon Colliders}
\author{I.~F.~Ginzburg\thanks{I am thankful for support Russian grants RFBR 08-02-00334-a and NSh-1027.2008.2 and INFN grant
}\\
  {\small  Sobolev Institute of Mathematics and
Novosibirk State University,  Novosibirsk, Russia,}
  }

 \maketitle

%%%%%%%%%%%%%%%%%%%%%%%%%%%%%%%%%%%%%%%%%%%%%%%%%%%%%%%%%%%%%%%%%%%%%
\begin{abstract}
In this report I discuss physical problems for future Photon Colliders (PLC), which can be stated  AFTER 10 years of work of LHC and few years of work of \epe ILC. I discuss mainly the unfavorable case when these colliders will give us only Higgs boson(s) and perhaps some charged particles of unclear nature. I focus my attention for the case of PLC based on the second stage of ILC (about 1 TeV) or CLIC (1-3 TeV). It offers opportunity to study new series of fundamental physical problems. Among them -- multiple production of gauge bosons, hunt for strong interaction in Higgs sector, search of exotic interactions in the process $\ggam\to \ggam$ with final photons having transverse momenta $\sim (0.5\div 0.7)E_e$.
\end{abstract}

%%%%%%%%%%%%%%%%%%%%%%%%%%%%%%%%%%%%%%%%%%%%%%
\section{Introduction. Different opportunities for PLC}
%%%%%%%%%%%%%%%%%%%%%%%%%%%%%%%%%%%%%%

We discuss here Photon Colliders (PLC) for different energy ranges. To do that, we start with
%\begin{figure}[hbt]
\begin{wrapfigure}
[9]{r}[0pt]{0.45\textwidth}\vspace{-5mm}
% \includegraphics[height=6.5cm,width=7.7cm]{volt.eps}
%\caption{}
 %  \label{figvolter}
% \end{wrapfigure}

\cl{\includegraphics[height=3.5cm,width=0.4\textwidth]{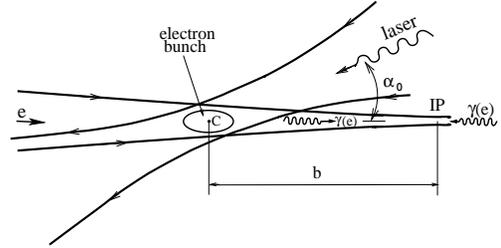}}
\caption{\it Photon Collider. Basic scheme}
  \label{figPLC}
 \end{wrapfigure}
repetition of basic scheme \cite{GKST}.
The focused laser flash meet the electron bunch of LC in the {\it
conversion point C} at small
distance $b$ before {\it
interaction point IP}. In { C} a laser photon scatters on high--energy electron taking from it a large portion of energy.
Scattered photons travel along the direction of the initial
electron with  angular spread $\sim 1/\gamma_e\equiv m_e c^2/E$, they are focused in the {IP}. Here they collide with opposite electron (\egam\ collider) or photon (\ggam\  collider).

For the ILC-1 based PLC,  the laser flash
with energy of a few Joules and length of a few mm is sufficient. The preferable form of basic electron beam for PLC is different from that for \epe LC. Based on that generally one can make the \ggam luminosity of PLC even larger than that of basic \epe LC. For discussed realizations this opportunity is used only weakly.  {\it The total additional cost is estimated in this case as $\sim 10$\% from that of LC} \cite{TESLA}.

\begin{figure}[hbt]
\includegraphics[width=0.24\textwidth,height=3cm]{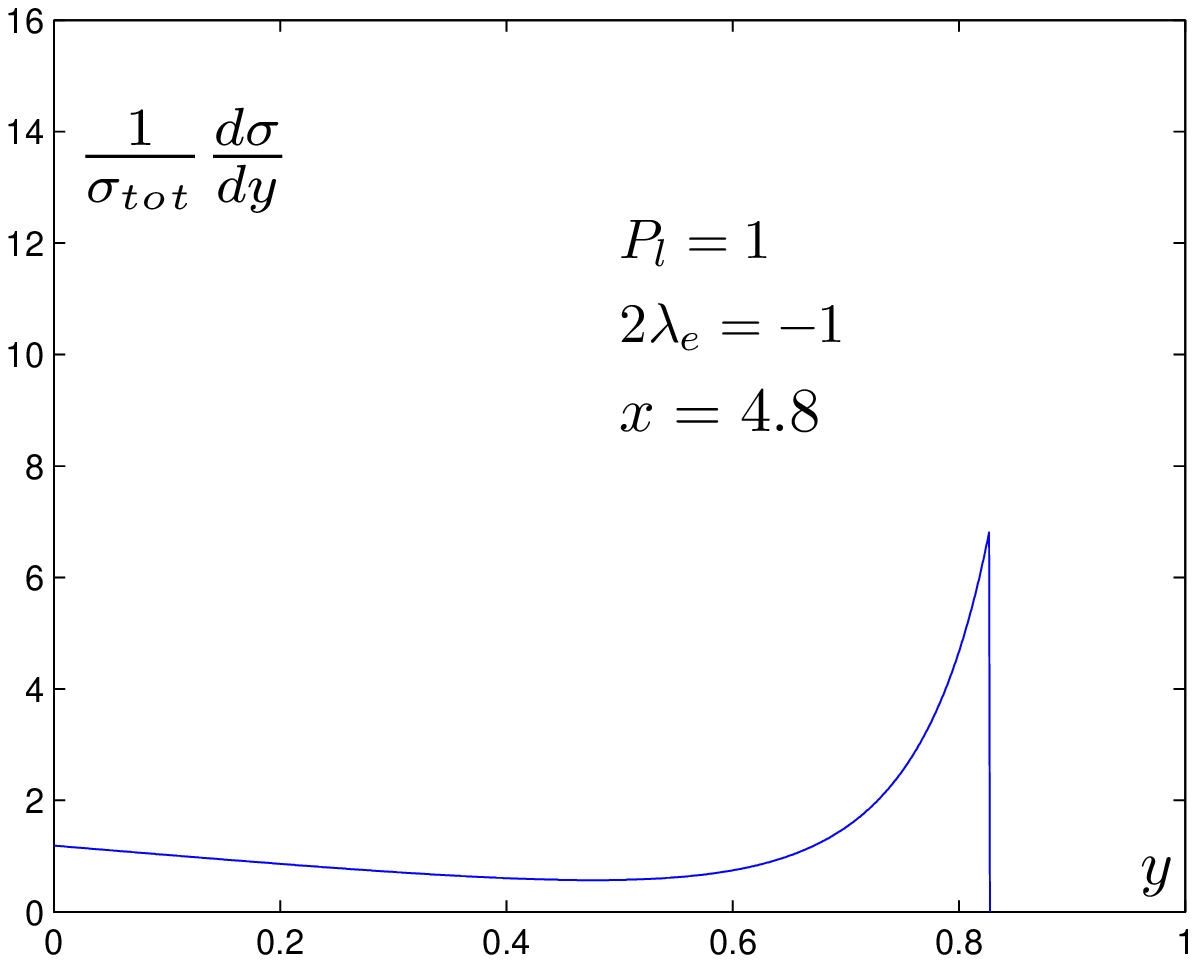}
\includegraphics[width=0.24\textwidth,height=3cm]{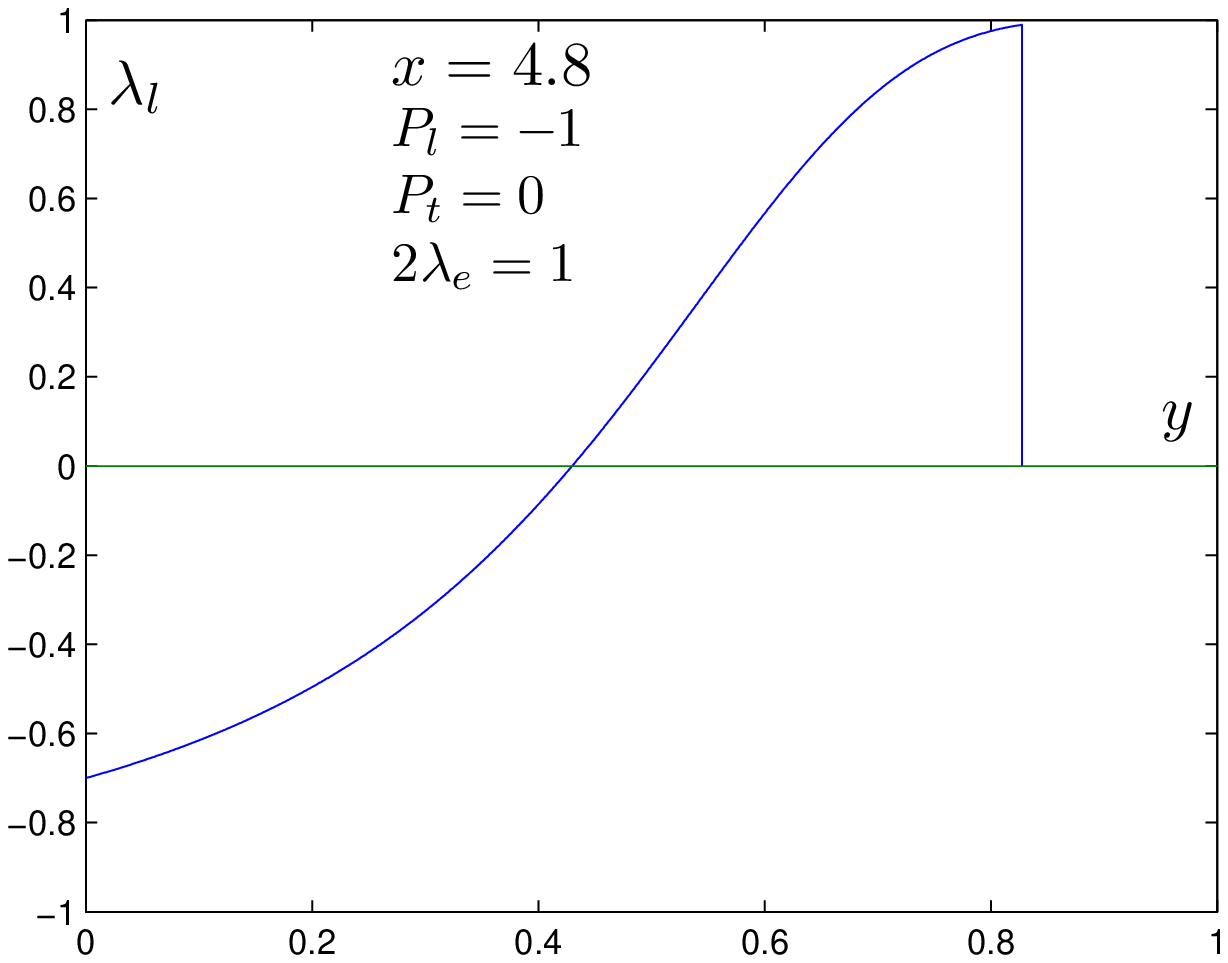}
\includegraphics[width=0.24\textwidth,height=3cm]{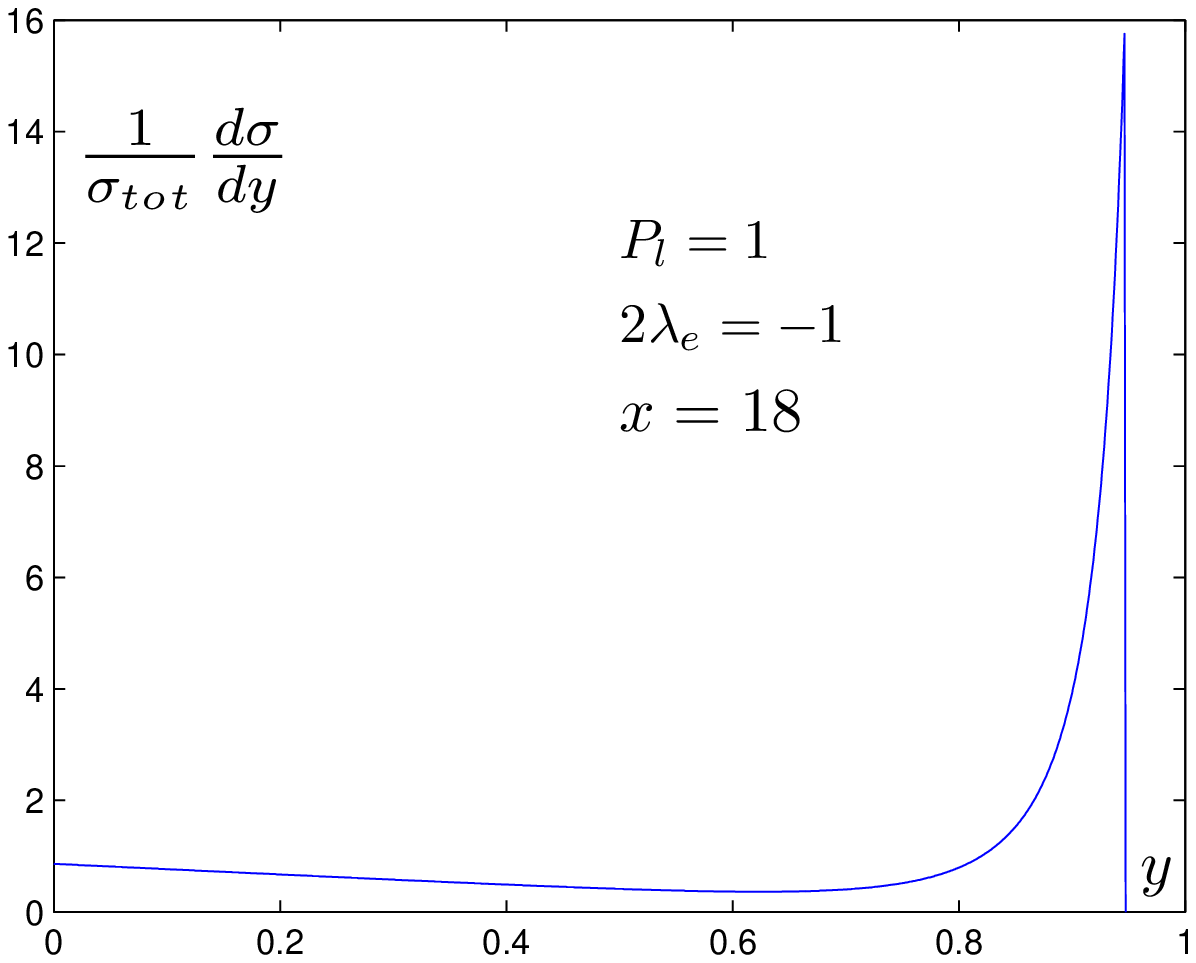}
\includegraphics[width=0.24\textwidth,height=3cm]{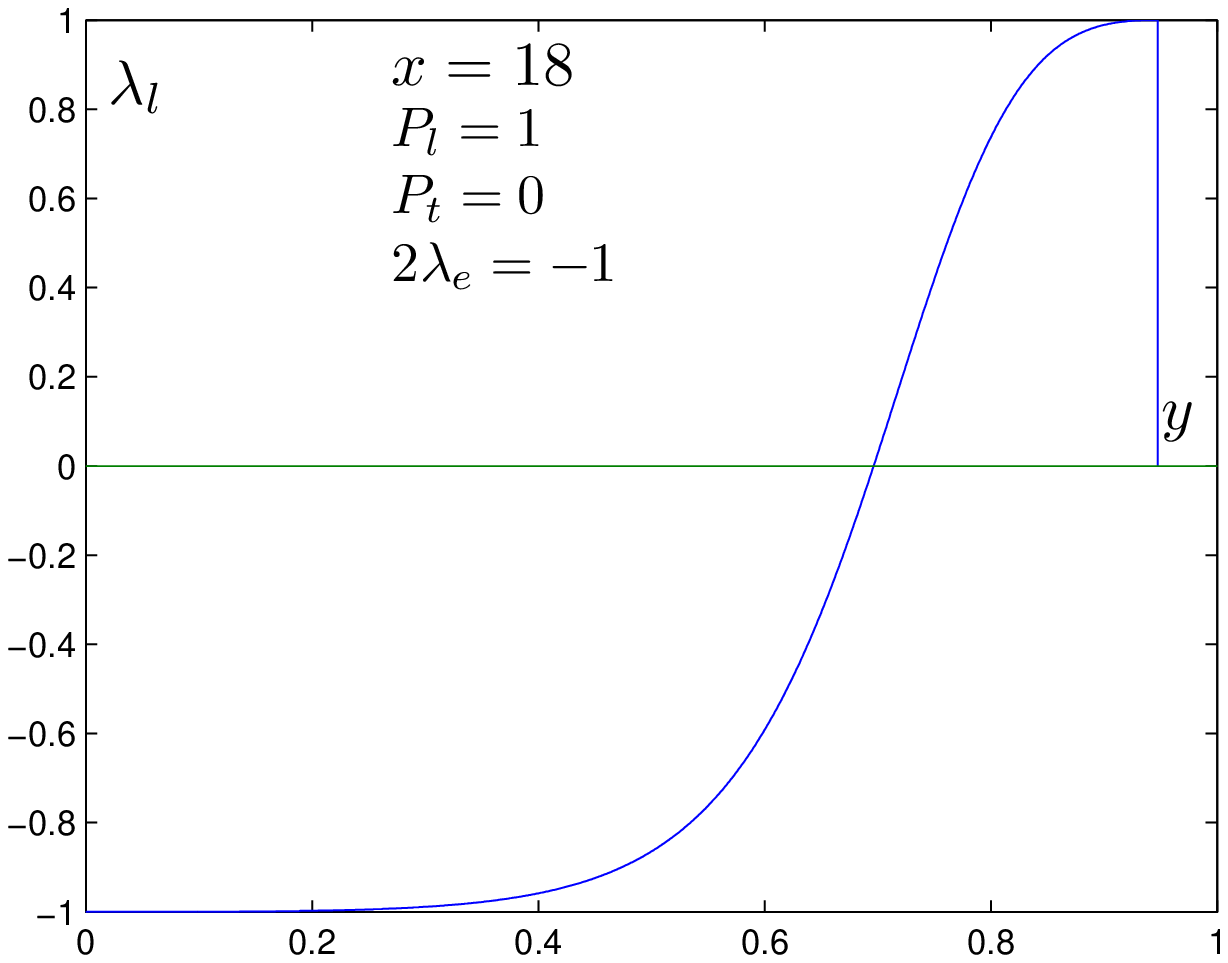}\\
\caption{\it Photon energy and polarization spectra, left $x=4.8$, right $x=18$. }
  \label{figphotspect}
 \end{figure}

The energy spectrum of obtained photon beam is concentrated
near its upper bound.  If $E_e$ -- electron energy and  $x=4E_e\omega_0/(m^2c^4)$, then $E_{\gamma,max}=E_e x/(x+1)$.
Spectrum become more sharp with suitable
choice of polarizations of initial electrons and laser
photons and with growth of $x$.
The obtained photon  beam is  strongly polarized.
The photon energy and mean photon helicity spectra are presented in Fig.~\ref{figphotspect} in dependence on $y=\omega/E_e$ for the case when initial electron helicity $\lambda_e=-1/2$ and initial lase photons are right polarized (helicity $P_l=1$) for two values $x$.

The real picture is more complex.\label{pagedeteriorate}\\ ({\it i}) When photons
with energy $\omega<\omega_{max}$ propagate from collision
point C to interaction point IP, they distribute over the
wider area reducing \ggam\ luminosity in its soft part.\\
({\it ii}) The low energy part of spectra is increased due
to multiple rescatterings of electron on the other laser
photons. \\
({\it iii}) The nonlinear QED effects also modify
 spectra, mainly for the case $x\le4.8$. \\
({\it iv}) At $x>4.8$ some fraction of produced photons disappear in the collision with laser photon, {$\gamma\gamma_0\to\epe$}. This effect result in strong limitation for the practical conversion coefficient.

{\it In future practice, the luminosity/polarization spectra must be measured
during operations.}

The production of photon beam for the LC with the electron  energy $E_e>250$~GeV offer difficult problems making construction of PLC for this energy range doubtful  \cite{Teln}. We consider here briefly two main ways of production of photon beams for $E_e\sim 1$~TeV \cite{GKS09}.\vspace{1.5mm}

{\bf The first way} is to use classical conversion scheme \cite{GKST} with infrared laser or FEL to reach the highest luminosity. The laser photon energy $\omega_0$ will be 0.5-0.2~eV with $x=4.8$ which prevents \epe pair production in collision of high energy and laser photons.  To get high conversion coefficient, the conversion process has to take place with large non-linear QED effects, making final photon distributions less monochromatic and less polarized. Here one must work with infrared optics which causes additional difficulties (see discussion e.g. in \cite{Teln}).\vspace{2mm}

{\bf The second way} is to use the same laser (and the same
optics) as for the electron beam energy 250~GeV (ILC-1) -- with photon energy  $\omega_0\sim 1$~eV -- but limit ourselves by a small conversion coefficient $k\le 0.14$ (at x=18) \cite{GKS09}. This value assures that the losses of high energy photons due to \epe pair production in collision of high energy photon with laser photon are small. At this value of conversion coefficient the non-linear QED effects are insignificant and contribution from rescatterings is small. Here the maximum photon energy is higher than in the first way, $\omega_m\approx (0.9-0.95)E$, energy distribution of high energy photons is more sharp, etc., right fig.~\ref{figphotspect}.  These advantages allow to consider this option despite the reduction of \ggam luminosity by about one order in comparison with the first way. {\it The second way seems more attractive to me.}\vspace{1.5mm}

The typical expected parameters of PLC for these two ways are presented in the Table. Here lines D-G describe only the high energy peak ($E_{\gamma 1,2}>0.7E_{\gamma max}$), which is separated well from
low energy part of spectrum and luminosity, it depends only weakly on details of conversion scheme. In both schemes one can hope to have annual luminosity $50\div 250$~fb$^{-1}$/year.
\begin{wraptable}{l}{0.7\textwidth}
\begin{tabular}{|c|l|c|c|}\hline
&\ \ \ \ \ \ \ \ \ Way $\to$&I, $x=4.8$ & II, $x=18$\\ \hline\hline
A&Necessary laser flash energy (J)&$<5$&$<5$\\\hline
B&The conversion coefficient $e\to\gamma$&0.7&0.15\\\hline
C&Maximal  photon energy $ E_{\gamma\,max}$&$0.8E_e$&$0.95E_e$\\\hline
D&Luminosity ${\cal L}_{\gamma\gamma}/{\cal L}_{e^+e^-}$& 0.35 & $0.03\div 0.05$\\\hline
E&Luminosity ${\cal L}_{e\gamma}/{\cal L}_{e^+e^-}$&0.25& 0.2\\\hline
F& Mean energy spread $<\Delta E_\gamma>$&0.07$E_{\gamma max}$&0.03$E_{\gamma max}$\\\hline
G&Mean photon helicity $<\lambda_\gamma> $&0.95&0.95\\\hline
\end{tabular}\vspace{2mm}
\label{Table1}
\caption{Parameters of PLC for two ways}
\end{wraptable}

The set of problems for PLC at ILC1 is widely discussed (see e.g. \cite{TESLA}). The  study of some of them (with
increase of thresholds for search of new particles) will be  a natural task for PLC with higher beam energy. We select here problems
to answer for questions: {\it what new can be studied at PLC  AFTER about 10 years of work of LHC with higher beam energy, and
perhaps, few years of work of \epe ILC with slightly larger beam energy and luminosity.}

%%%%%%%%%%%%%%%%%%%%%%%%%%%%%%%%%%%%%%%%%%
\section{QCD and hadron physics}
%%%%%%%%%%%%%%%%%%%%%%%%%%%%%%%%%%%%%%%%%%%%%%%%%

{\bf Photon structure function} is unique object of QCD, calculable at large enough $Q^2$ without additional phenomenological parameters \cite{Witten}. It can be measured at PLC in \egam\ mode with high accuracy, since photon target with its energy and polarization here are practically known. The manipulation with beam polarizations will be important instrument here.

The region of electron transverse  momenta above 50~GeV~($M_Z/2$)  can be studied well, providing opportunity to study  effect of {\bf $\pmb Z$-boson exchange and $\pmb {\gamma^*-Z}$ interference.}

The other studies like those at HERA are possible here.

%%%%%%%%%%%%%%%%%%%%%%%%%%%%%%
\section{Higgs physics}
%%%%%%%%%%%%%%%%%%%%%%%%%%%%%%%%%

Higgs mechanism of EWSB can be realized either by minimal Higgs sector with one observable neutral scalar Higgs boson (SM) or  by non-minimal Higgs sector with larger number of observable scalars. In this section for definiteness we  consider SM and specific non-minimal Higgs sector -- Two Higgs Doublet Model (2HDM). The latter is the simplest extension of Higgs sector of SM.  It contains 2 complex Higgs doublet fields $\phi_1$ and $\phi_2$ with v.e.v.'s $v\cos\beta$ and  $v\sin\beta$. The physical sector contains charged scalars  $H^\pm$ and three neutral scalars $h_i$, generally having  no definite \CP parity. In the CP conserving case
these three $h_i$ become  two scalars $h$, $H$
($M_h<M_H$) and a pseudoscalar $A$. For definiteness, we assume
the Model II for the Yukawa coupling in 2HDM (the same is
realized in MSSM).

%%%%%%%%%%%%%%%%%%%%%%%%%%%%%%%%%%%%
\paragraph{SM-like scenario. Distinguishing models.}
Let earlier observations discover Higgs boson, similar to that in SM ({\it  SM-like scenario}).
{How to state whether we deal with SM Higgs boson or some other realization of Higgs sector (e.g. 2HDM)? What can we say about properties of this realization?}

LHC can measure Higgs couplings to particles only with  low precision,
typically 10-20\%.
The \epe LC will improve these results up to 5-10\%, sometimes better.
The PLC can improve these accuracies further to about 2\%.

Here, measuring the\ $h\ggam$ ($hZ\gamma$) couplings is very
promising. The expected accuracy in the measurement of the
two-photon width is  2\% at $M_h\le 150$ GeV and
$\int{\cal L}dt= 30$~fb$^{-1}$ (by 5 times lower than the
anticipated annual luminosity) \cite{Jik}.

{\bf Example -- distinguishing SM/2HDM.} The SM -- like scenario means that the
\begin{wrapfigure}[12]{l}[0pt]{0.4\textwidth}\vspace{-7mm}
\includegraphics[height=4.3cm,width=0.35\textwidth]{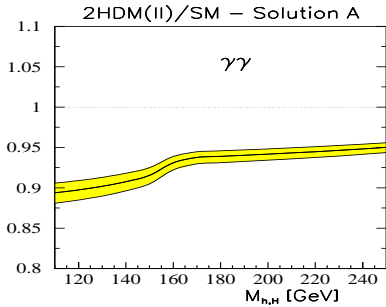}\vspace{-4mm}
\caption{{\it The ratio of
$\Gamma(h \to \ggam)$  to its SM value for typical class of realization of SM-like scenario.
}}
  \label{HSMlike}
 \end{wrapfigure}
coupling constants
{\it squared},  measured at LHC and \epe LC, are {\it  close} to the SM value within anticipated precision, (not coupling
constants themselves!. In the 2HDM this scenario can be
realized in many ways.

The models can be distinguished via measurement of
the \ggam\ width of the observed SM-like
Higgs boson, fig.~\ref{HSMlike} \cite{GKO}. In this figure we show the ratio of $\Gamma(h \to \ggam)$  to its SM value for one typical class of realization of SM-like scenario.
The bands reflect the anticipated uncertainty of future measurements.
The deviation from $ SM$, given by contributions
of heavy charged Higgs bosons for a  natural set of
parameters, is about $ 10$\% (compare with anticipated
2\% accuracy). For the other sets of parameters, consistent with SM-like scenario, the deviation from $SM$ is even larger.

\paragraph{CP violation in Higgs sector.}
In many  extensions of Higgs model (e.g. in 2HDM)  observable neutral Higgs bosons $h_i$ have generally {\bf no} definite CP-parity and effectively
 \be
{\cal L}_{\gamma\gamma H} = G_\gamma^{SM}\left[g_\gamma
HF^{\mu\nu}F_{\mu\nu} + i \tilde{g}_\gamma
HF^{\mu\nu}\tilde{F}_{\mu\nu}\right];\;\;g_\gamma\sim \tilde{g}_\gamma\sim 1\,.\label{Higgsano}
 \ee
Here $F^{\mu\nu}$  and $\tilde
{F}^{\mu\nu}=\varepsilon^{\mu\nu\alpha\beta}F_{\alpha\beta}/2$
are the standard field
strength for the electromagnetic  field.
The relative effective couplings $g$ and $\tilde{g}$ are
described with standard triangle diagram $H\ggam$, they are
expressed with known equations via masses of charged fermions and W, and mixing parameters (parameters of 2HDM potential). They are generally complex ($b\bar{b}$ --loop).

Total production cross section varies strong with variation of circular { $\lambda_i$} and linear { $\ell_i$} polarizations of photon beams and  the angle { $\psi$} between linear polarization vectors  \cite{GIv}:
\bear{c}
\sigma(\ggam\to H)=\sigma^{SM}_{np}%\times
%\\[3mm]
\times\left[|g_\gamma|^2
 (1+\lambda_1\lambda_2
 +\ell_1\ell_2 \cos 2\psi)
 +|\tilde{g}_\gamma|^2
 \left(1+\lambda_1\lambda_2-\ell_1\ell_2 \cos 2\psi \right)+\right.\\[3mm] +
\left. 2Re(g_\gamma^*\tilde{g}_\gamma)(\lambda_1+\lambda_2)
 +2Im(g_\gamma^*\tilde{g}_\gamma)
 \ell_1\ell_2\sin 2\psi\right]\,.
 \eear{HiggsCPgen}
\begin{wrapfigure}[10]{l}[0pt]{0.48\textwidth}\vspace{-5mm}
\includegraphics[height=4cm,width=0.45\textwidth]{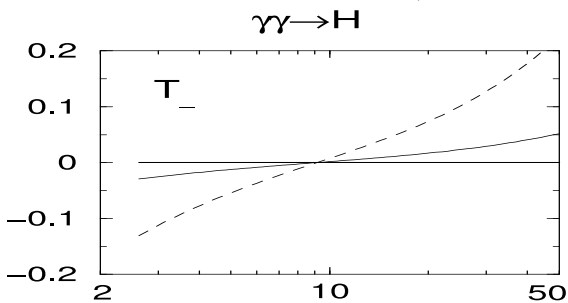}

\caption{\it Effect of CP violation in 2HDM.}

\label{hCPasym}
\end{wrapfigure}

In particular, violation of CP symmetry in the Higgs sector leads to difference in the $\ggam\to H$ production cross sections in the collision of photons with identical total helicity (0) but with opposite helicities of separate photons:
\bear{c}
\!\! T_-\!=\!\fr{\sigma(\lambda_i)\!-\!\sigma(-\lambda_i)}{\sigma_{np}^{SM}}\!\propto\!
(\lambda_1\!+\!\lambda_2)Re(g_\gamma\tilde{g}^*_\gamma).
\eear{HiggsCPas}

Standard calculation of vertexes in the 2HDM at different parameters of model gives typical dependence, shown in fig.~\ref{hCPasym}  at $\lambda_1=\lambda_2=\pm 1$. It is seen that effect is strong and can be measured well.

%%%%%%%%%%%%%%%%%%%%%%%%%%%%%%%%%%%%%%%%%%%%%%%%%%%%%%%%%%%%%%%%%
\paragraph{Observation of strong interaction in Higgs sector  in $e\gamma\to eWW$ process at not too high energy.}
At high values of Higgs boson self-coupling
constant, the Higgs mechanism of Electroweak Symmetry Breaking in
Standard Model (SM) can be realized without actual Higgs boson but
with strong interaction in Higgs sector (SIHS) which will manifest
itself as a strong interaction of longitudinal components of $W$ and
$Z$ bosons. It is expected that this interaction will be  seen in the form of $W_LW_l$, $W_LZ_L$ and $Z_LZ_L$
resonances at $1.5\div 2$~TeV. Main efforts to discover this opportunity are directed
towards the observation of such resonant states. It is a difficult task for
the LHC due to high background and it cannot be realized at the
energies reachable at the ILC in its initial stages.

This strong interaction can be observed in the study of the  charge asymmetry of produced $W^\pm$ in the process $e^-\gamma\to e^-W^+W^-$ similar to that which was discussed in low energy pion physics \cite{eepipi}, \cite{SD}. To explain the set up of the problem we discuss  this process in SM \cite{GKstrong}.

We subdivide the diagrams of the process into three groups, where subprocesses of main interest are shown in boxes, sign $\otimes$ represents next stage of process.

a) Diagrams  $e^-\to e^-\gamma^*(Z^*)\otimes \boxed{\gamma\gamma^*(\gamma Z^*)\to W^+W^-}$ contain subprocesses
$\gamma\gamma^*\to W^+W^-$ and $\gamma Z^*\to W^+W^-$, modified
by the strong interaction in the Higgs sector ({\it two--gauge
}).

b) Diagrams  $\gamma e^-\!\to\! e^{-*}\!\to\! e^-\gamma^*(Z^*)\!\otimes\!\boxed{\gamma^*(Z^*)\!\to\! W^+W^-}$ contain subprocesses $\gamma^*\!\to\!
W^+W^-$ and $ Z^*\to W^+W^-$, modified by the strong interaction
in the Higgs sector  ({\it one--gauge}).

c) Diagrams $\gamma\oplus\boxed{e^-\to W^-W^+e^-}$
are prepared by connecting
the photon line to each charged particle line
to the diagram shown inside the box. Strong
interaction does not modify this contribution. These contributions are switched off at suitable electron polarization.

The subprocess  $\gamma\gamma^*\to W^+W^-$ (from contribution a)) produces C-even system $W^+W^-$, the subprocess  $\gamma^*\to W^+W^-$ (from contribution b)) produces C-odd system $W^+W^-$. The interference of similar contributions for the production of pions is responsible for large enough charge asymmetry, very sensitive to the phase difference of S (D) and P waves in $\pi\pi$ scattering, \cite{eepipi}. This very phenomenon  also takes place in the discussed case of $W$'s. However, for the production of $W^\pm$ subprocesses with the replacement of  $\gamma^*\to Z^*$ are also essential. Therefore, the final states of each type  have no definite $C$-parity. Hence, charge asymmetry appears both due to interference between contributions of types a) and b) and  due to interference of $\gamma^*$ and $Z^*$ contributions  each within
their own types.

{\bf Asymmetries in SM.}
To observe the main features of the effect of charge asymmetry and its potential for the study of strong interaction in the Higgs sector, we
calculated some quantities describing charge asymmetry
for $e^-\gamma$
collision at $\sqrt{s}=500$~GeV with polarized photons. We
used  CalcHEP package \cite{CompHEP} for simulation.

We denote by  $p^\pm$  momenta of  $W^\pm$, by
$p_e$ -- momentum of the scattered electron and $
w=\fr{\sqrt{(p^++p^-)^2}}{2M_W}$,
  $v_1=\fr{\la(p^+-p^-)p_e\ra}{\la(p^++p^-)p_e\ra}$.
We present below dependence of charge asymmetric quantity $v_1$ on $w$.  The $w$-dependencies for the other charge asymmetric quantities have similar qualitative features \cite{GKstrong}.

We applied {\it the cut in transverse momentum of the scattered
electron,}
 \be
p_\bot^e\ge p_{\bot 0}\;\;\mbox{ with
}\,\;a)\; p_{\bot 0}=10~GeV,\,\,b)\; p_{\bot 0}=30~GeV.
           \ee
Observation of the scattered electron allows to check kinematics completely.

\begin{figure}[hbt]
\begin{tabular}{ccc}
  \includegraphics[width=0.32\textwidth, height=4cm]{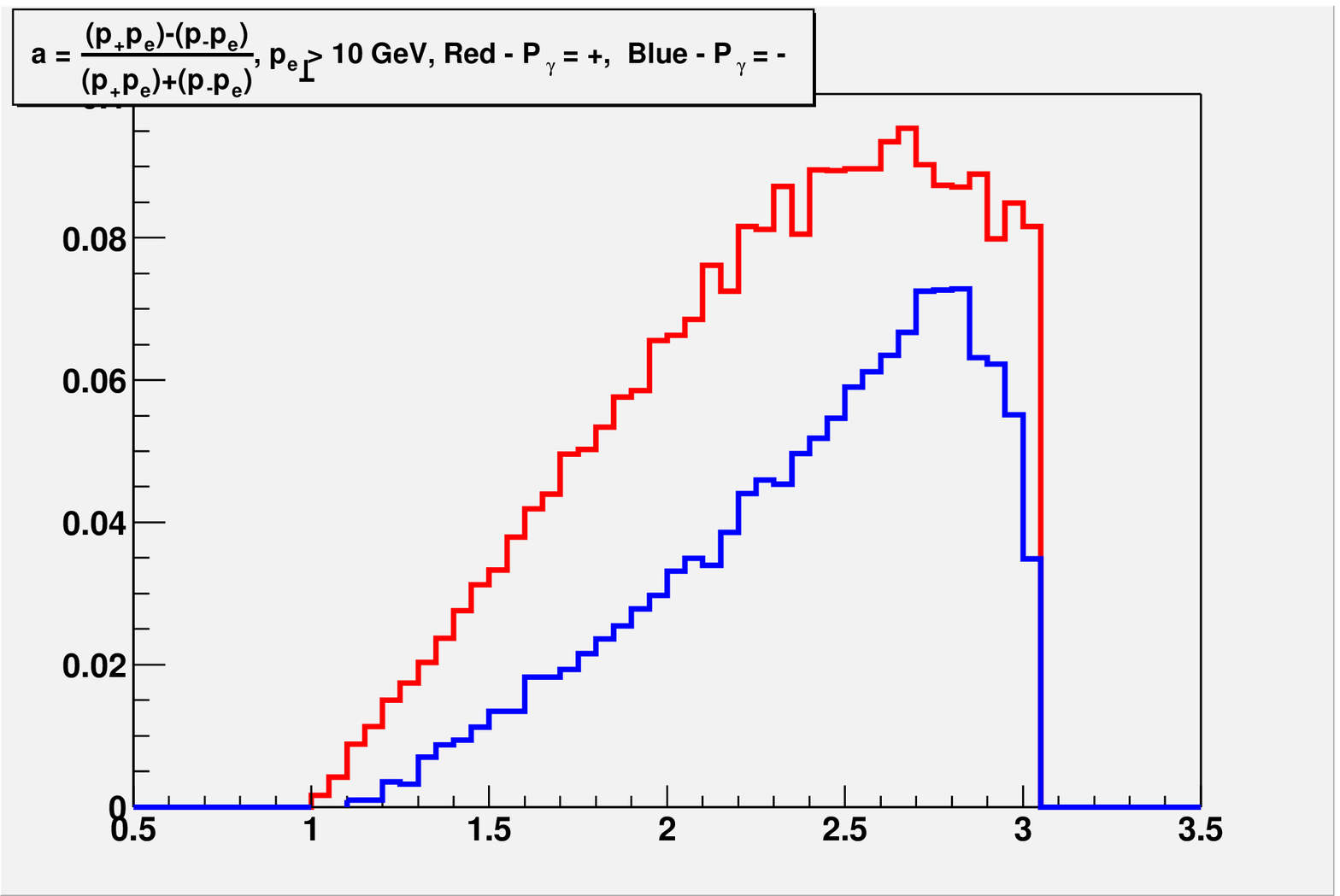}&
  \includegraphics[width=0.32\textwidth, height=4cm]{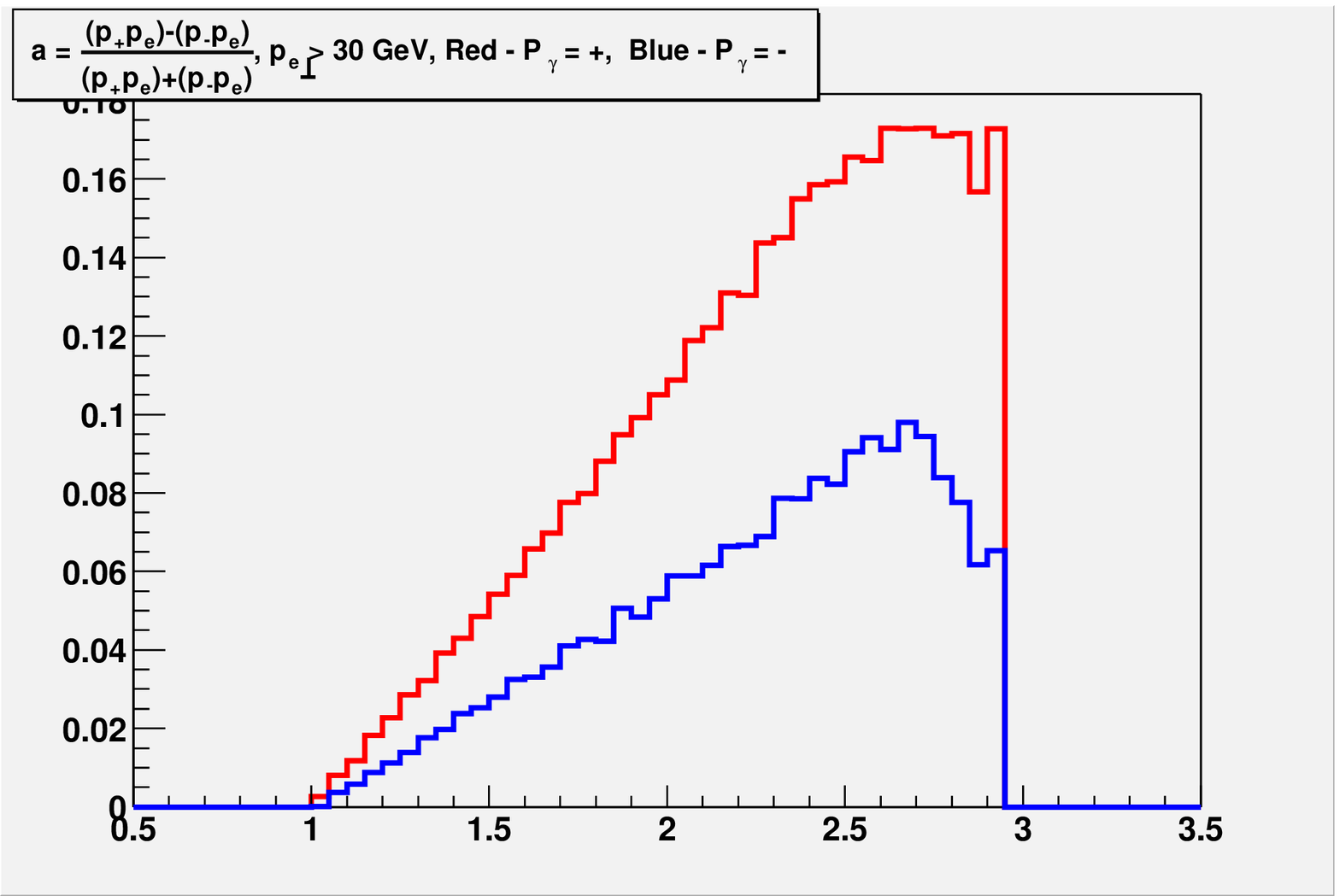}&
  \includegraphics[width=0.32\textwidth,height=4cm]{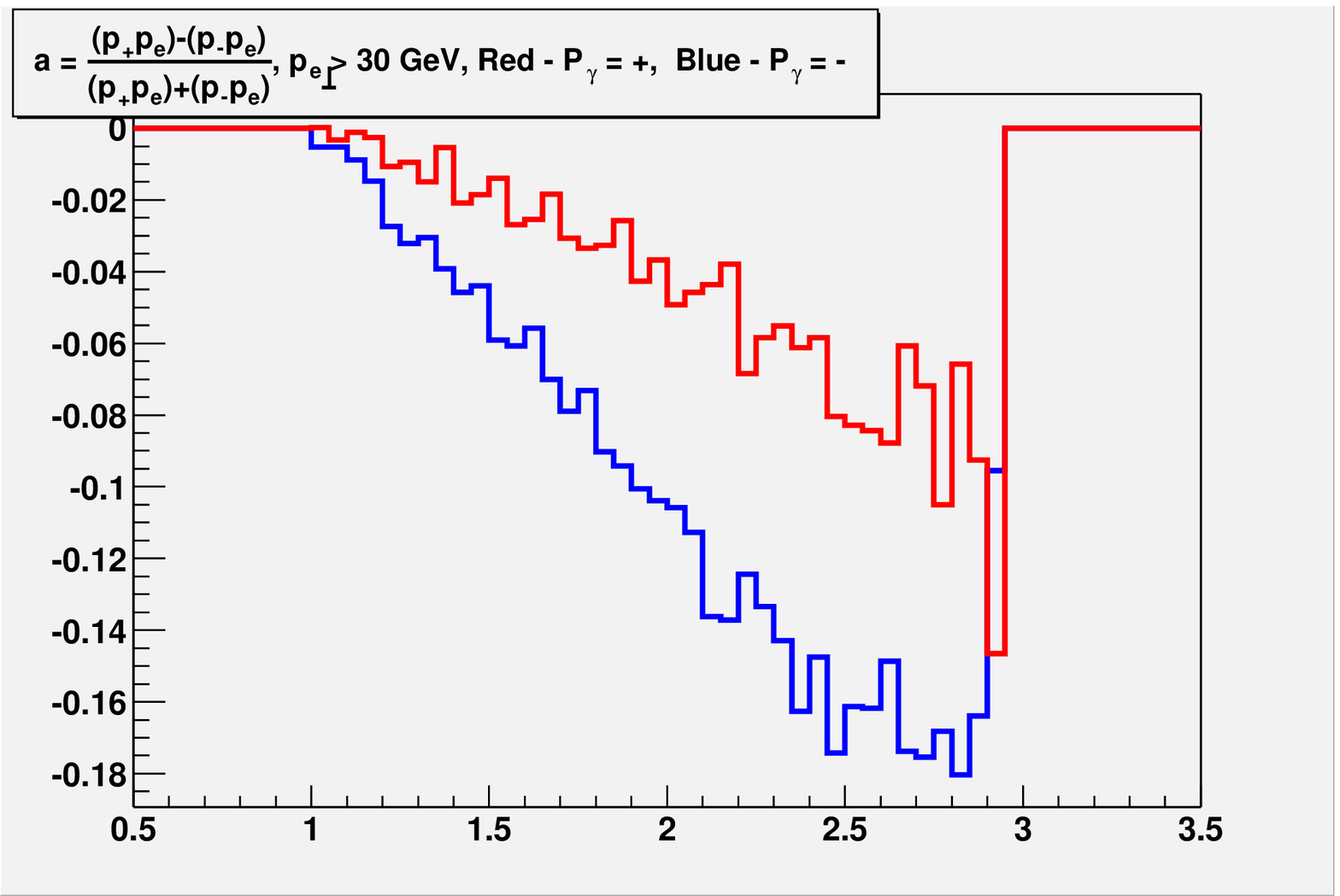}\\
 $p_{\bot0}=10$~GeV&$p_{\bot0}=30$~GeV&$p_{\bot0}=30$~GeV, without\\&& one-gauge contributions
\end{tabular}

  \caption{\it Distribution of $v_1$ in dependence on $w$.
  The upper curves are for right-hand polarized photons, the lower curves are for left-hand polarized photons
    }
  \label{Ginzburg_fig2}
\end{figure}

{\it  Influence of polarization}.  Fig.~\ref{Ginzburg_fig2} (left and central plots) represents distribution in  variable $v_1$ on photon polarization and cut in $p_\bot^e$.
We did not study the dependence on electron polarization.
This dependence is expected to be weak in SM where main
contribution to cross section is given by diagrams of type
a) with virtual photons having the lowest possible
energy. These photons "forget" the polarization of the incident
electron. The strong interaction contribution becomes
essential at highest effective masses of $WW$ system with
high energy of virtual photon or $Z$, the helicity of which
reproduces almost completely the helicity of incident
electron \cite{GSerb}. The study of this
dependence will be a necessary part of studies beyond SM.

{\it Significance of different contributions.}
To understand the extent of the effect of interest, we compared
the entire distribution in variable $v_1$ with that without one-gauge
contribution at $p_{\bot 0}=30$~GeV (right plot in Fig.~\ref{Ginzburg_fig2}).
Strong interaction in the Higgs sector modifies both one--gauge
and two--gauge contributions. The study of charge
asymmetry caused by their interference will be a source of information on  this
strong interaction. One can see that one--gauge contribution is so essential that neglecting on it  even changes the sign of charge asymmetry (compared to that for the entire process).
Therefore, the charge asymmetry is very sensitive to the interference of two--gauge and one--gauge contributions which is modified under the strong interaction in the Higgs sector. The measurement of this asymmetry will be a source of data on the phase difference of different partial waves of $W_LW_L$ scattering.

 \paragraph{If  more than one scalar, like Higgs boson, will be observed,}
it will be strong argument in favor of more complex Higgs sector, like 2HDM or something else. It is necessary to measure properties of these scalars, including coupling to fermions, gauge bosons and self-couplings with the best accuracy, to find what model is realized.

To understand properties of model, one must first to measure masses all scalars and their coupling to gauge bosons and some fermions. However, even these data are non-sufficient for fixing of parameters of model. Usually for this goal somebody suggest to measure triple Higgs coupling in the processes like $\epe\to Zhh$, $\ggam\to hh$. However their cross sections are typically low and contributions of triple Higgs vertexes there is added by contributions of product of other Higgs vertexes. Moreover, knowledge of this vertex is non-sufficient for fixing of parameters of the model. It was found in \cite{Kan} the  complete set of observable parameters of 2HDM can be extracted from {\it masses of $H^\pm$ and 3 neutrals $h_1$, $h_2$, $h_3$} (generally with no definite CP parity) and {charged Higgs $H^\pm$}, {\it their couplings to gauge bosons}, added by {\it 3 triple Higgs couplings} (like $h_ih_ih_i$ or  $H^+H^-h_i$) and {\it one quartic coupling} (like $H^+H^-H^+H^-$). At high enough energy of PLC the cross sections of processes $\ggam\to H^+H^-h_i$ are $\propto \lambda_{H^+H^-h_i}$ without interference withe other vertexes and they are not small. One can hope also to measure coupling $\lambda_{H^+H^-H^+H^-}$ via measuring of production $\ggam\to H^+H^-H^+H^-$ cross section.

The information of the complete set of parameters of model will give also information about way of evolution of phase states of earlier Universe \cite{GIK}.

%%%%%%%%%%%%%%%%%%%%%%%%%%%%%%%%%%
\section{New particles}
%%%%%%%%%%%%%%%%%%%%%%%%%%%%%

New charged particles will be discovered at LHC and in \epe mode of LC. We expect their decay for final states with invisible particles (like LSP in MSSM).\\
\bu { How to measure mass, decay modes and spin of these new particles?}\\
In these problems the $\gamma\gamma$ production provides essential advantages compared to $e^+e^-$ collisions\\
\bu {How to observe signals from new neutral particles -- possible candidates for dark matter?}

 The cross section of the  {\it pair production}
$\gamma\gamma\to P^+P^-$ ($P=S$ -- scalar, $P=F$ -- fermion, $P=W$
-- gauge boson) not far from the threshold is given by QED
with reasonable accuracy.
\begin{figure}[htb]
\includegraphics[height=4.5cm,width=0.99\textwidth]{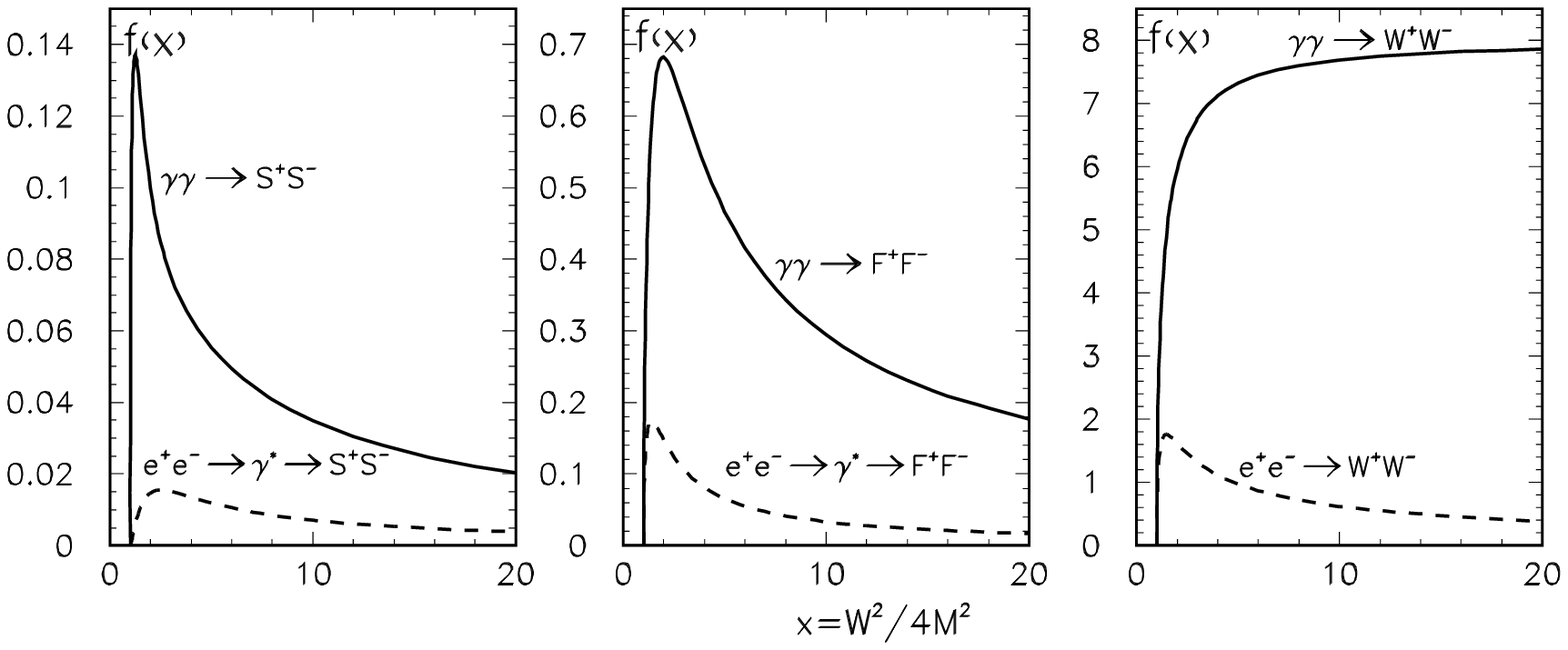}

\caption{\it $\fr{\sigma(\gamma\gamma\to P^+P^-)}{\pi\alpha^2/M_P^2}$,
nonpolarized photons, and $\fr{\sigma(e^+e^-\to
\gamma^*\to P^+P^-)}{\pi\alpha^2/M_P^2}$ }
\label{pairQED}
\end{figure}
$\bullet$ {These  cross sections decreases slowly with
energy growth}. Therefore, they can be studied
relatively far from the threshold where the decay
products are almost non-overlapping.

\bu Near the threshold {$f_P\propto (1
+\lambda_1\lambda_2 \pm\ell_1\ell_2 \cos2\phi)$}\  with + sign for $P=S$  and -- sign for $P=F$. This polarization dependence provides the opportunity to determine spin of produced particle $P$ in the experiments with longitudinally polarized photons.

\bu The polarization of produced fermion or vector $P$ depends   on the initial photon helicity. At the $P$ decay this polarization
is transformed into the momentum distribution of decay products.
E.g., for the SM processes like {$\ggam\to
\mu^+\mu^- +neutrals$} (obtained from muon decay modes of
$\ggam\to WW$, $\ggam\to\tau^+\tau^-$, etc.) muons should
exhibit {charge asymmetry linked to the polarization of
initial photons} -- see sect.~\ref{secchas}.
These studies can help to understand the {\it nature of candidates for Dark Matter particles.}

The possible CP\ violation  in the $P\gamma$ interaction can be seen as a variation of cross section with changing the sign of both photon helicities (like in fig.~\ref{hCPasym}).

\paragraph{Charge asymmetry in processes
                    $\pmb{\gamma_\uparrow\gamma_\uparrow \to \mu^+\mu^-\nu_\mu\bar{\nu}_\mu}$,
            $\pmb{\gamma_\uparrow\gamma_\uparrow\to W^\pm\mu^\mp\nu}.$ \label{secchas}}

In the SM the  effect appears due to P nonconservation  in the W-decay. 

We select events with two cuts, for escape angle $\theta$
for each observed particle and for transverse momentum of each observed particle and for missed transverse momentum
 \be
 \pi-\theta_0 > \theta > \theta_0\,,\qquad
 p_\bot > p_{\bot \mu}^c\,.\label{cutschas}
 \ee
These simultaneous cuts allow to eliminate many backgrounds.
We used $\theta_0=10$~mrad and study $ p_{\bot \mu}^c$ dependence of effect starting from $ p_{\bot \mu}^c=10$~GeV.

\begin{figure}[htb]
\includegraphics[width=0.97\textwidth,height=4.5cm]{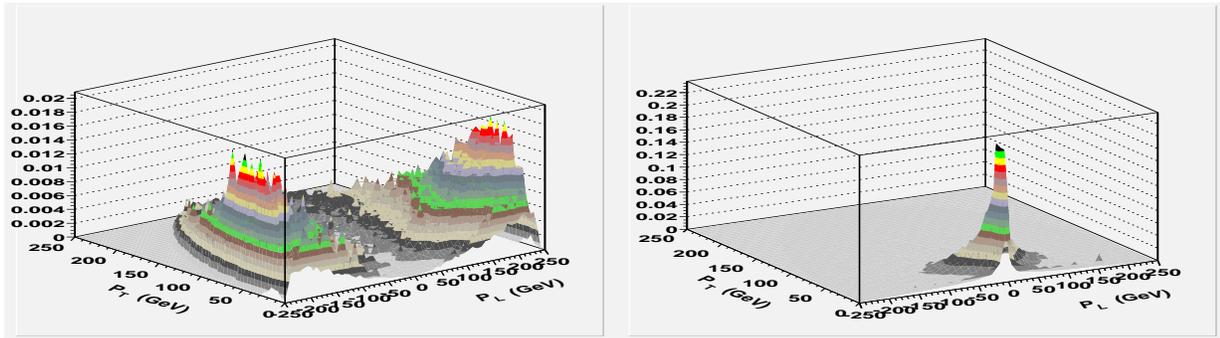}
\caption{\it Difference between distributions of positive and negative
        muons for $\gamma_{-}\gamma_{-}\to W\mu\nu$}
\label{figchasgen}
\end{figure}

The fig.~\ref{figchasgen} demonstrates the charge asymmetry in the collision of two left polarized photons at $\sqrt{s}=500$~GeV, left and right plots show $p_\bot,\,p_L $ distributions for negative and positive muons respectively. We find that effect is strong and well observable even at large enough $p_{\bot \mu}^c=100$~GeV (the results were obtained  with the aid of CalcHEP  package \cite{AGKPP}). So, one can conclude that this charge asymmetry is huge and well observable effect in SM. The study of $p_{\bot \mu}^c$ dependence of effect shows that one can hope to see effects of New Physics in these asymmetries at high transverse  momenta (larger than 100 GeV).

%%%%%%%%%%%%%%%%%%%%%%%%%%%%%%%%%%%%%%%%%%%%%%%%%%%%%%%%%%
\section{Multiple production of SM gauge bosons}
%%%%%%%%%%%%%%%%%%%%%%%%%%%%%%%%

The observation of pure interactions of SM gauge bosons (W and Z) or their interaction with leptons will allow to check SM with higher accuracy and observe signals of New Physics.
The most ambitious  goal is to find deviations from predictions
of SM caused by New Physics
\begin{wrapfigure}[13]{l}[0pt]{7cm}\vspace{-3mm}
 \includegraphics[width=0.4\textwidth,height=5cm]{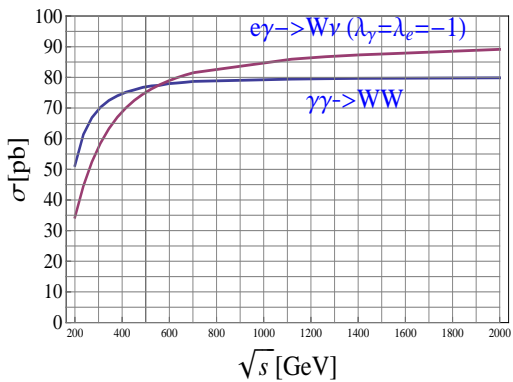}\vspace{-3mm}
\caption{\it Cross sections of 2-nd order processes}
  \label{fig2WW}
 \end{wrapfigure}
interactions (and described by anomalies in effective Lagrangian). There are many anomalies relevant to the gauge boson interactions. Each process is sensitive to some group of
anomalies. Large variety of processes obtainable at
PLC's allows to separate anomalies from each
other.
The high energy PLC is the only collider among different future accelerators where one can measure large number of different processes of such type with high enough accuracy.\\
$\left.\;\;\;\right.${\bf 2-nd order processes.} The cross sections of basic processes \ggww and \gewnu are so high
(Fig.~\ref{fig2WW}) that one can expect to obtain about $10^7$ events per year providing accuracy better than 0.1\%.
The cross sections are almost independent of energy and
photon polarization
\cite{GKPS}. However, final  distributions  depend on polarization strongly \cite{AGKPP}.

The accuracy of measurement of these cross sections is sufficient to study in detail 2-loop radiative corrections. Together with standard problems of precise calculations one can note here two non-trivial problems, demanding detailed theoretical study:\\
{\it{\it(i)}  construction of
 $S$--matrix for system with unstable particles;\\
{\it(ii)} gluon corrections like Pomeron exchange between
quark components of $W$'s.}

The mentioned high values of  cross sections of the 2-nd order processes make it possible to measure their multiple "radiative derivatives" --- processes of the 3-rd
and 4-th order, depending in different ways on various anomalous contributions to the effective Lagrangian.

\begin{figure}[htb]
\centering
\begin{tabular}{cc}
\includegraphics[height=0.22\textheight,width=0.45\textwidth]{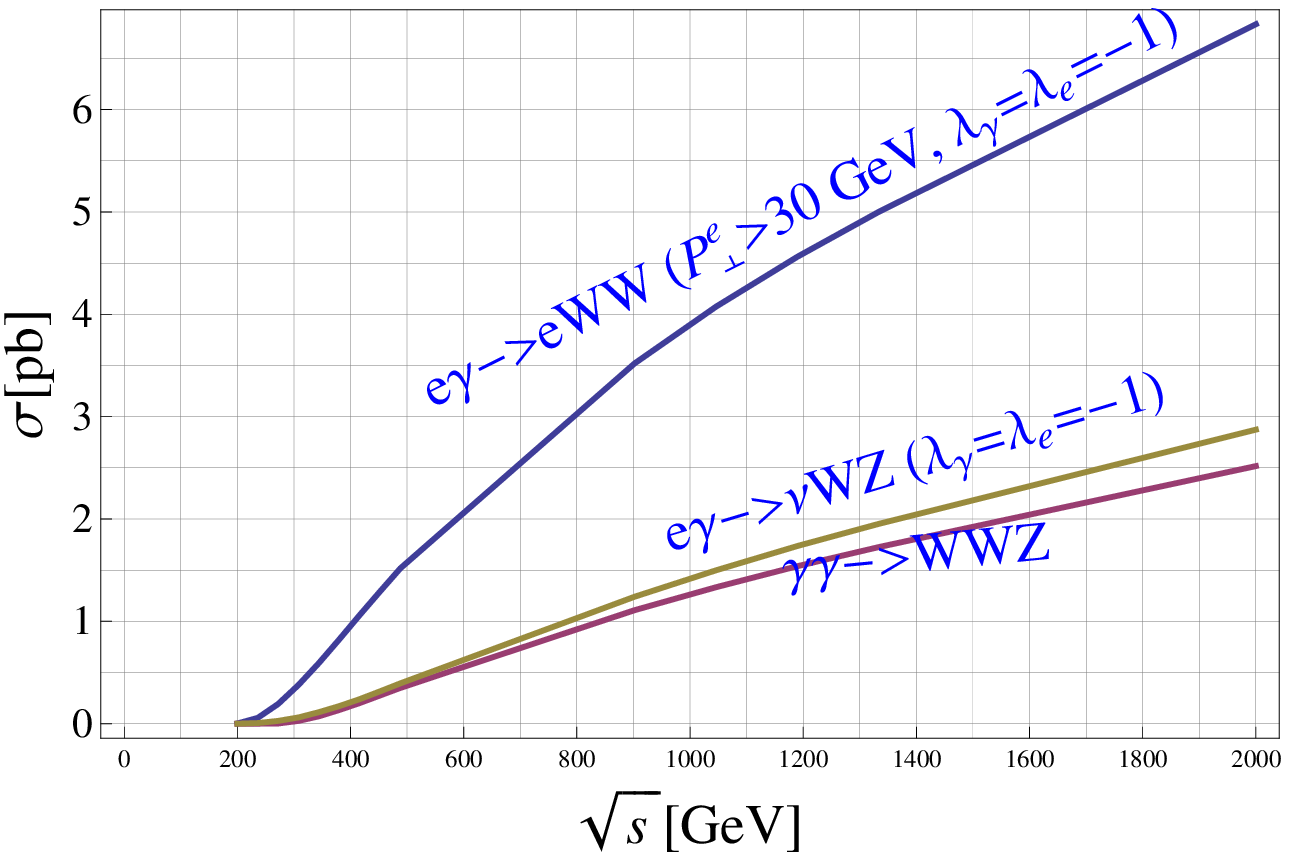}&
\includegraphics[height=0.22\textheight,width=0.45\textwidth]{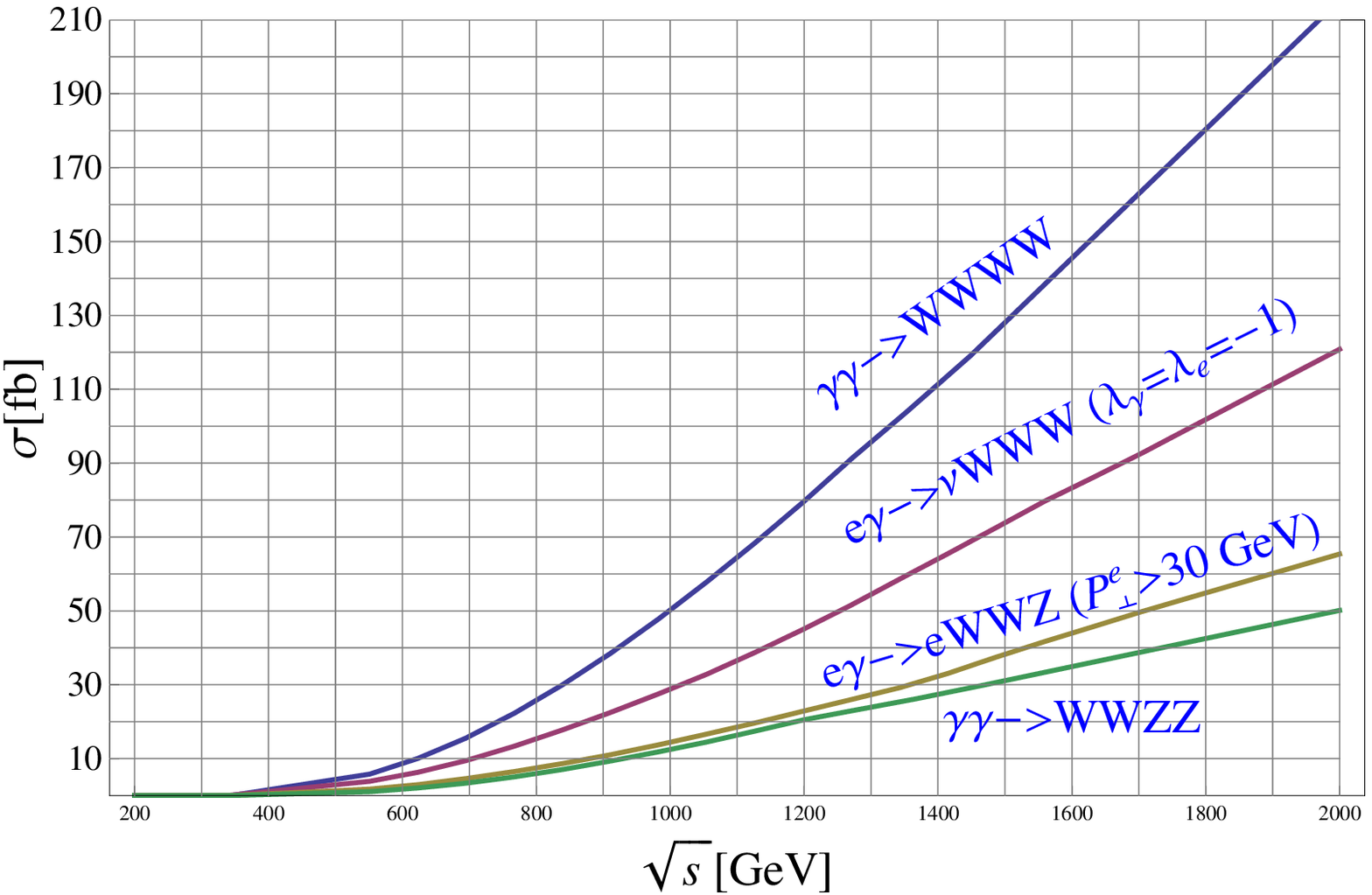}\\
{\bf(a)} 3-rd order processes & {\bf(b)} 4-th order processes
\end{tabular}
\caption{\it  Cross sections of  3-rd  and   4-th order processes}
  \label{fig3WW}
\end{figure}

{\bf 3-rd order processes.}  We consider here 3 processes (fig.~\ref{fig3WW}a).
Total cross section
$\sigma_{e\gamma\to eWW}\simeq
dn_\gamma\otimes \sigma_{\gamma\gamma\to WW}$. It is very high and easily estimated by equivalent photon method. This
large contribution is not very interesting, being only a cross section of \ggww averaged with some weight. However, at large enough
transverse momentum of scattered electron this factorization is
violated. Because of it we present $\sigma_{e\gamma\to eWW}$ only for  $p_{\bot e}>30$~GeV. Even this small fraction of total cross section appears so large  that  it allows {\it to separate contribution of $\gamma Z\to WW$ subprocess}.

{\bf 4-th order processes.} The cross sections of these processes (Fig.~\ref{fig3WW}b) are high enough to
measure them with 1\% precision. For the same reason as for process $e\gamma\to eWW$
we present cross section for process $e\gamma\to eZWW$ only for  $p_{\bot e}>30$~GeV. Even this small fraction of total cross section appears so large  that  it allows {\it to separate contribution of $\gamma Z\to WWZ$ subprocess}.

The study of the 2-nd order processes will allow to extract some anomalous parameters or their combinations. The study of the 3-rd order processes will allow to enlarge the number of extracted anomalous parameters and separate some of combinations extracted from the 2-nd order processes. The study of the 4-th order processes will again enlarge the number of separated anomalous parameters.

%%%%%%%%%%%%%%%%%%%%%%%%%%%%%%%%%%%%%%%%%%%%%%%%%%%%%
\section{Large angle high energy photons for exotics}
%%%%%%%%%%%%%%%%%%%%%%%%%%%%%%%%%%%%%%%%%%%%%%%%%%%%%%
The PLC allows to observe signals from the whole group of exotic models of New Physics in one common experiment. These are models with {\it large extra dimensions} \cite{LED}, {\it point-like monopole} \cite{monop}, {\it  unparticles} \cite{unpart}.
All these models have common signature -- the  cross section for $\ggam\to\gamma\gamma$ production grows with energy as $\omega^6$ ($\omega=\sqrt{s}/2$) and
the photons are produced
\begin{wrapfigure}{r}{0.2\textwidth}%\vspace{-4mm}
\cl{\includegraphics[height=1.8cm,width=0.18\textwidth]{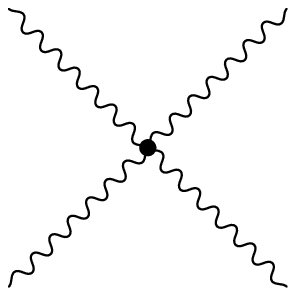}}%\vspace{-3mm}
 \caption{\it Effective Lagrangian}
 \label{exoticdiagr}
 \end{wrapfigure}
almost isotropically. Future observations either will give limits for scales of these exotics or will allow to see these effects by recording large $p_\bot\sim (0.5\div 0.7) E_e$ photons\fn{\it In my personal opinion it is hardly probable that these models describe reality.}. The study of dependence on initial photon polarization will be useful to separate the mechanisms.

All these exotics at modern day energies can be described by  effective point-like interaction of Fig.~\ref{exoticdiagr}:
\be
L\propto \fr{ F^{\mu\nu}F^{\alpha\beta}F_{\rho\sigma}F_{\phi\tau}}{\Lambda^4},\quad({ \Lambda^2\gg s/4}).\label{exoticLagr}
 \ee
In different models different orders of field indices are realized, $\Lambda$ is characteristic mass scale, expressed via parameters of model. (In all cases  $s$, $t$ and $u$ -- channels are essential.)

Let us describe  main  features of matrix element (in the photon c.m.s.):\\
\bu{gauge invariance provides  factor $\omega$ for each photon
leg;}\\
\bu  to make this factor dimensionless it should be written   as
$\omega/\Lambda$. Therefore,  the amplitude ${\cal M}\propto (\omega/\Lambda)^4=s^2/(2\Lambda)^4$.\\
The  characteristic scale $\Lambda$ is large enough  not to contradict modern day data. It accumulates other  coefficients.
The cross section
 \be
\sigma_{tot} =\fr{1}{32 \pi s}
\left(\fr{s}{4 \Lambda^2}\right)^4\,,\quad d\sigma=
\sigma_{tot}\Phi\left(\fr{p_\bot^2}{s}\right)\,\fr{2dp_\bot^2}{\sqrt{s(s-4p_\bot^2)\,}}.
\label{TOTCREX}
 \ee
\begin{wraptable}{r}{0.48\textwidth}\vspace{-6mm}
\centerline{\begin{tabular}{|l|c|c|}\hline
&$\Lambda$&reference\\\hline
Tevatron D0& 175~GeV&\cite{D0}\\\hline
LHC & 2 TeV&\cite{monop}\\\hline
$\ggam$ (100 fb$^{-1}$)& $3E_e$&\cite{monop}\\\hline
$\epe$ LC (1000 fb$^{-1}$)& $2E_e$&\cite{monop}\\\hline
\end{tabular}}
\caption{\it The obtainable discovery limits.}
\label{tab:limits}
\end{wraptable}
with smooth  function $\Phi(p_\bot^2/s)$, describing some composition of S and P-waves, dependent on details of model,   and
 $
 \int \Phi(z)\,\fr{2dz}{\sqrt{1-4z\,}}=1\,. $
For large extra dimensions and monopoles entire $s$ dependence is given by  the factor $s^4/(2\Lambda)^8$ from \eqref{TOTCREX}, for unparticles additional factor $(s/4\Lambda^2)^{d_u-2}$ is added.

For the {\bf large extra dimensions} case the point in Fig.~\ref{exoticdiagr} describes an elementary interaction, given by product of stress-energy tensors $T_{ab}$ for the incident and the final photons, that are exchanging the tower of Kaluza-Klein excitations (with permutations), i.e. $
{\cal M}_{\ggam\to\ggam}\propto \left\la T_{ab}T^{ab}/\Lambda^4
\right\ra\approx F^{\mu\nu}F_{\nu\alpha}F^{\alpha\beta}F_{\beta\mu}/\Lambda^4\;\;+\;\;{ permutations}$.
After averaging over polarizations for tensorial KK excitations
  \be
\Phi\propto 2\left(1-p_\bot^2/{s}\right)^2=(3+\cos^2\theta)^2/8=
(\hat{s}^4+\hat{t}^4+\hat{u}^4)/2\hat{s}^4\,.\label{distrLED}
 \ee

Unlike to ILC1, at high energy PLC the other channels (like  $\ggam\to WW$) are less sensitive to the extra dimension effect.

{\bf The point--like Dirac monopole} existence would explain mysterious
quantization of an electric charge since in this case $ge=2\pi n$ with $n=1,\,2,...$. {\it There is no place
for this monoplle in modern theories of our world but there are no
precise reasons against its  existence.} In this case the point in Fig.~\ref{exoticdiagr} corresponds to exchange of loop of heavy monopoles (like electron loop in QED -- Heisenberg--Euler type lagrangian).

Let $M$ be monopole mass. At $s\ll M^2$ the electrodynamics
of monopoles is expected to be similar to the standard QED with effective perturbation parameter { $g\sqrt{s}/(4\pi M)$} \cite{monop}. The $\ggam\to\ggam$ scattering is described by monopole loop, and
it is calculated within QED,
 $$
 {\cal L}_{4\gamma}=
 \fr{1}{36} \left(\fr{g}{\sqrt{4\pi} M}\right)^4
\left[\fr{\beta_++\beta_-}{2}
\left(F^{\mu\nu}F_{\mu\nu}\right)^2
+\fr{\beta_+-\beta_-}{2}\left(F^{\mu\nu}\tilde{F}_{\mu\nu}\right)^2\right]\,.
$$
The coefficients $\beta_\pm$ and details of angular and polarization dependence depend strongly on the spin of the monopole.

After averaging over polarizations, the $p_\bot$ dependence and total cross section is  described by the same equations as for the extra dimensions case. The  parameter $\Lambda$ is expressed via monopole mass and coefficient  $a_J$, dependent on monopole spin $J$ ($n=1,\,2,...$):
 \be
\Lambda=(M/n)a_J,\;\;\; where\;\;\; a_0=0.177,\;\; a_{1/2}=0.125,\;\; a_1=0.069.
\ee

{\bf  Unparticle }  $\cal U$ is an object, describing particle scattering via propagator which has no poles at real axis. It was introduced in 2007  \cite{unpart}. This propagator behaves (in the scalar case) as $(-p^2)^{d_U-2}$ where scalar dimension $d_u$  is not integer or half-integer. The interaction carried by unparticle is described as $\fr{F^{\mu\nu}F_{\mu\nu}{\cal{U}}}{\Lambda^{2d_U}}$ with some phase factor. For matrix element it gives
  \bear{c}
 {{\cal M}=\fr{ F^{\mu\nu}F_{\mu\nu}F^{\rho\tau}F_{\rho\tau}}{\Lambda^{4d_U}}(-P^2)^{d_U-2}\;\; +}\;\;  { permutations}\,.\\[3mm]
  |{\cal M}|^2=C\fr{s^{2d_U}+|t|^{2d_U}+|u|^{2d_U}+\cos(d_u\pi)[(s|t|)^{d_U}+(s|u)^{d_U}]
+(tu)^{d_U}}
{\Lambda^{4d_U}}
\eear{}

{\bf The anticipated discovery limits} for all these models are shown in the Table~\ref{tab:limits}. The results of D0 experiment \cite{D0}, recalculated to used notations, are also included here. For the unparticle model presented numbers are modified by corrections  $\propto(d_U- 2)$.\\

This paper is supported by grants RFBR 08-02-00334-a, NSh-1027.2008.2,
Program of Dept. of Phys. Sc RAS "Experimental and theoretical studies
of fundamental interactions related to LHC" and INFN grant.

\end{document}